# Use of a Golden Gate plasmid set enabling scarless MoClo-compatible transcription unit assembly

Stijn T. de Vries[1], Laura Kley[1] and Daniel Schindler[1,2,*]

[1] Max Planck Institute for Terrestrial Microbiology, Karl-von-Frisch-Str. 10, D-35043 Marburg, Germany

[2] Center for Synthetic Microbiology (SYNMIKRO), Philipps-University Marburg, Karl-von-Frisch-Str. 14, D-35032 Marburg, Germany

* Correspondence to: daniel.schindler@mpi-marburg.mpg.de

**Abstract**

Golden Gate cloning has become a powerful and widely used DNA assembly method. Its modular nature and the reusability of standardized parts allow rapid construction of transcription units and multi-gene constructs. Importantly, its modular structure makes it compatible with laboratory automation, allowing for systematic and highly complex DNA assembly. Golden Gate cloning relies on Type IIS enzymes that cleave an adjacent undefined sequence motif at a defined distance from the directed enzyme recognition motif. This feature has been used to define hierarchical Golden Gate assembly standards with defined overhangs ("fusion sites") for defined part libraries. The simplest Golden Gate standard would consist of three part libraries, namely promoter, coding and terminator sequences, respectively. Each library would have defined fusion sites, allowing a hierarchical Golden Gate assembly to generate transcription units. Typically, Type IIS enzymes are used, which generate four nucleotide overhangs. This results in small scar sequences in hierarchical DNA assemblies, which can affect the functionality of transcription units. However, there are enzymes that generate three nucleotide overhangs, such as SapI. Here we provide a step-by-step protocol on how to use SapI to assemble transcription units using the start and stop codon for scarless transcription unit assembly. The protocol also provides guidance on how to perform multi-gene Golden Gate assemblies with the resulting transcription units using the Modular Cloning standard. The transcription units expressing fluorophores are used as an example.

# 1. Introduction

Golden Gate cloning and the resulting Modular Cloning (MoClo) has become a popular DNA assembly strategy because of its modular and hierarchical structure, which allows for the economical construction of complex DNA assemblies [1,2]. Golden Gate cloning is based on the use of Type IIS restriction enzymes, which recognize a specific, directed DNA sequence but cut any adjacent DNA sequence at a defined distance (Fig. 1A). The Type IIS enzymes used for Golden Gate assembly generate sticky ends. These sticky ends can be defined by the user, allowing the creation of part libraries, e.g. promoter, coding (CDS) and terminator sequences, each defined position with a specific Type IIS-generated overhang ("fusion site") (Fig. 1B). The basic parts are usually cloned into defined level 0 plasmids, sequence validated and can be released with a Type IIS enzyme to allow the assembly of individual parts into transcription units (TU).

This modular structure allows for standardized workflows, making this DNA assembly strategy particularly amenable to automation. Another advantage of this standardized method is that once the basic parts have been sequence validated, they are no longer amplified, eliminating the risk of PCR-based introduction of sequence errors. Therefore, only basic parts need to be sequence validated, TUs and higher order assemblies can be validated by simple, rapid and economic diagnostic methods such as colony PCR (cPCR) or restriction pattern analysis. We have recently developed protocols for semi-automated Golden Gate DNA assembly and long-read sequencing-based construct validation that allow economic DNA assembly reactions in 1 µL total volume and low-cost high-throughput sequence validation [3-6]. Establishing a systematic Golden Gate assembly standard with corresponding part libraries can therefore be highly rewarding, as it allows rapid combinatorial DNA assembly to answer biological questions or test the application potential of the gene(s) of interest.

The most interesting part of MoClo, however, is that it theoretically allows the assembly of DNA constructs of infinite size (Fig. 1C). The trick of the MoClo system is that TUs can be assembled into multiple hierarchical level 1 plasmids, each corresponding to a defined position in higher order multi-gene DNA constructs. The level 1 TU plasmids differ only in the fusion sites created when the fragments are released with the appropriate Type IIS endonuclease. This allows for the assembly of multiple level 1 plasmids into a higher order DNA assembly. With the original MoClo strategy, up to 6 TUs can be assembled into a level M plasmid in combination with the appropriate end linker (EL) [2]. Again, there are multiple level M plasmids, similar to level 1 plasmids. Level M multi-gene constructs can be released with another Type IIS enzyme and allow the assembly of multiple level M constructs into a level P plasmids. There are multiple level P plasmids, similar to level 1 and level M. The clue is that at each level, the enzyme and antibiotic selection marker are changed and the level P DNA assemblies are again compatible with level M-based Golden Gate cloning, allowing for an infinite cloning cycle. This strategy has been used to generate highly complex DNA constructs for various applications such as the generation of complex repetitive sequence arrays from sequence libraries or the construction of synthetic chromosomes [7,8].

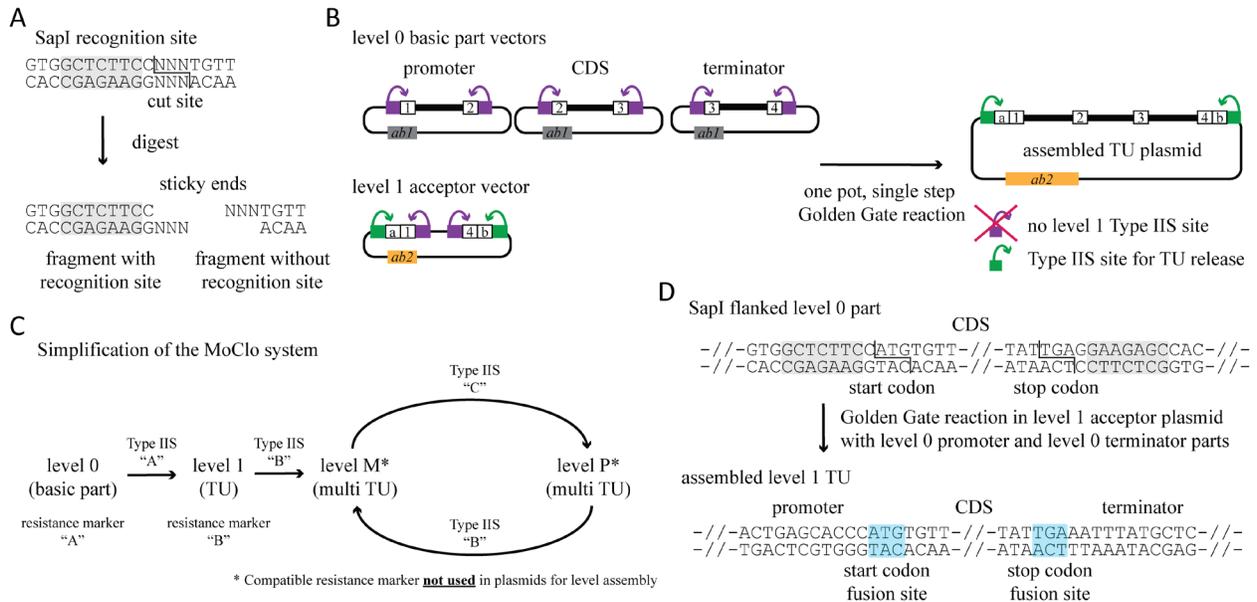

**Figure 1 | Basic principles of Golden Gate cloning and MoClo. (A)** Type IIS enzymes recognize a directed DNA sequence and cut any sequence at a defined distance from the recognition site. This results in two DNA fragments, one with and one without the recognition site. SapI is used as an example and its recognition site is shaded gray. N indicates that any DNA sequence can be at the cut site (note: enzyme activities can be inhibited by DNA methylation). **(B)** Golden Gate cloning relies on the orchestration of Type IIS enzyme recognition and cleavage sites. The use of defined fusion sites (here represented by 1 to 4) allows the assembly of multiple parts into an acceptor plasmid. The basic part plasmids and the acceptor plasmid differ at least in the selection marker and the orientation of the Type IIS enzyme recognition sites. The assembled plasmid does not contain the Type IIS recognition site used to generate the plasmid. However, a different Type IIS enzyme recognition site may be present in the acceptor plasmid to release the assembled DNA fragment with new fusion sites (here indicated by a and b) for higher order assembly. **(C)** The Modular Cloning (MoClo) system allows in theory for an infinite assembly rounds of DNA fragments [2]. The generated basic parts (level 0) can be assembled in transcription units (TU). Multiple TUs can be assembled into higher order DNA assemblies. By swapping the Type IIS enzyme and the antibiotic selection marker between each step, in particular between level M and level P allows in theory for unlimited cycles to combine DNA assemblies of growing complexities. The only requirement is careful planning to ensure that the fragments have the appropriate fusion sites. Important: In this protocol level 1 plasmids are used that are compatible with level P. The procedure is identical except for the Type IIS enzyme and the antibiotic resistance markers. For simplicity, the standard MoClo system is visualized and the corresponding level M compatible plasmids are also available. **(D)** Assembly of transcription units based on three nucleotide overhangs using the start codon (ATG) and stop codon (TGA) as fusion sites to allow scarless TU construction.

However, Golden Gate cloning has a small disadvantage. When using defined part libraries, the fusion sites are fixed. Mostly Type IIS enzymes are used, which result in four nucleotide fusion sites. This results in small scars at the fusion sites, which can affect TU functionality. Especially when using experimental systems that are sensitive to small changes, such as the construction and application of small synthetic regulatory RNAs (sRNAs), scar sequences introduced by DNA assembly should be avoided [3,9]. To minimize scar sequences introduced at the level of TUs, the use of SapI seems to be promising. In contrast to the commonly used Type IIS enzymes, SapI has a three nucleotide fusion site, which allows the use of codons as fusion sites [10]. Recently, cloning fidelities have been determined for several Type IIS enzymes, allowing for data-driven, optimized Golden Gate assembly strategies, including the SapI enzyme [11].

The protocol outlined describes the use of SapI for the construction of TUs, resulting in no assembly scars based on the three nucleotide sticky ends, in contrast to the commonly used four nucleotide sticky end-generating Type IIS enzymes (Fig. 1D). The use of SapI allows for codon-based cloning, using the start (ATG) and stop (TGA) codons to link the CDS to the appropriate promoter and terminator sequences. The protocol describes the design and generation of basic parts (3.1) for their SapI-mediated assembly into TUs (3.2) and subsequent higher order assemblies (3.3). The assembled TUs (3.2) are compatible with the standard MoClo system developed by Sylvestre Marillonnet's group. The protocol concludes with an exemplary evaluation of the assembled multi-gene construct using spotting assays to test microbial growth and to measure the resulting fluorescence of the assembled multi-gene constructs (3.4).

## 2.    Materials

The conduction of the described protocol requires the following standard laboratory equipment and consumables.

1. Standard reagents, consumables and instrumentation for PCR reactions.
2. Standard reagents and equipment for gel electrophoresis.
3. Standard reagents, consumables and instrumentation for microbial culturing.
4. Standard reagents, consumables and instrumentation for transformation of *E. coli*.
5. Standard reagents, consumables and instrumentation for plasmid extraction and DNA clean up (*see* **Note 1**).
6. Standard micropipettes and consumables, 12 channel pipettes are advised.
7. Equipment to visualize distinct fluorescence signals on colony level (e.g., imaging system or binocular). Alternatively, a plate reader may be used.

2.1 Plasmids
All plasmids relevant for this protocol are listed in Table 1. Plasmids can be requested from the corresponding author.

**Tab. 1 | Plasmids used in this study**

| Name | Relevant features | Reference |
|---|---|---|
| pSL099 | Level 0 acceptor plasmid for promoter position[*], spectinomycin resistance marker, Type IIS enzyme for level 0 assembly: BpiI Type IIS enzyme for fragment release: SapI | unpublished |
| pSL102 | Level 0 acceptor plasmid for CDS position[*], spectinomycin resistance marker, Type IIS enzyme for level 0 assembly: BpiI Type IIS enzyme for fragment release: SapI | unpublished |
| pSL106 | Level 0 acceptor plasmid for terminator position[*], spectinomycin resistance marker, Type IIS enzyme for level 0 assembly: BpiI Type IIS enzyme for fragment release: SapI | unpublished |
| pSL108 | Level 1 TU acceptor plasmid position 1, ampicillin resistance marker, Type IIS enzyme for level 1 assembly: SapI Type IIS enzyme for fragment release: BsaI | unpublished |
| pSL109 | Level 1 TU acceptor plasmid position 2, ampicillin resistance marker, Type IIS enzyme for level 1 assembly: SapI Type IIS enzyme for fragment release: BsaI | unpublished |
| pSL110 | Level 1 TU acceptor plasmid position 3, ampicillin resistance marker, Type IIS enzyme for level 1 assembly: SapI Type IIS enzyme for fragment release: BsaI | unpublished |
| pMA67 | Level P multi-gene acceptor plasmid , kanamycin resistance marker, Type IIS enzyme for level P assembly: BsaI Type IIS enzyme for fragment release: BpiI | [7] |
| pMA676 | Level P Endlinker plasmid for multi-gene assembly, chloramphenicol resistance marker, Type IIS enzyme for level P assembly: BsaI | [12] |
| pSL842 | Level 0 basic CDS part *mTurquoise*, spectinomycin resistance marker, Type IIS enzyme for fragment release: SapI | unpublished |
| pSL843 | Level 0 basic CDS part *gfp*, spectinomycin resistance marker, Type IIS enzyme for fragment release: SapI | unpublished |
| pSL206 | Level 0 basic CDS part *mCherry*, spectinomycin resistance marker, | unpublished |

| Name | Relevant features | Reference |
| --- | --- | --- |
| | Type IIS enzyme for fragment release: SapI | |
| pSLcol_16.03 | Level 0 basic promoter part medium strength, spectinomycin resistance marker, Type IIS enzyme for fragment release: SapI | unpublished |
| pSL209 | Level 0 basic terminator part T1-T7Te, spectinomycin resistance marker, Type IIS enzyme for fragment release: SapI | unpublished |
| pSL844 | Level 1 position 1 plasmid containing neutral linker, ampicillin resistance marker, Type IIS enzyme for fragment release: BsaI | unpublished |
| pSL845 | Level 1 pos. 1 *mTurquoise* TU, ampicillin resistance marker, Type IIS enzyme for fragment release: BsaI | unpublished |
| pSL846 | Level 1 position 2 plasmid containing neutral, ampicillin resistance marker, Type IIS enzyme for fragment release: BsaI | unpublished |
| pSL847 | Level 1 pos. 2 *gfp* TU, ampicillin resistance marker, Type IIS enzyme for fragment release: BsaI | unpublished |
| pSL848 | Level 1 position 3 plasmid containing neutral, ampicillin resistance marker, Type IIS enzyme for fragment release: BsaI | unpublished |
| pSL849 | Level 1 pos. 3 *mCherry* TU, ampicillin resistance marker, Type IIS enzyme for fragment release: BsaI | unpublished |
| pSL765 | Level P multi-gene assembly with m*Turquoise*, *gfp*, and *mCherry* TU, kanamycin resistance marker, Type IIS enzyme for fragment release: BpiI | unpublished |
| pSL766 | Level P multi-gene assembly with m*Turquoise* TU, *gfp* TU, and neutral linker, kanamycin resistance marker, Type IIS enzyme for fragment release: BpiI | unpublished |
| pSL767 | Level P multi-gene assembly with m*Turquoise* TU, neutral linker, and *mCherry* TU, kanamycin resistance marker, Type IIS enzyme for fragment release: BpiI | unpublished |
| pSL768 | Level P multi-gene assembly with neutral linker, *gfp* TU, and *mCherry* TU, kanamycin resistance marker, Type IIS enzyme for fragment release: BpiI | unpublished |
| pSL769 | Level P multi-gene assembly with neutral linker at all three positions, kanamycin resistance marker, Type IIS enzyme for fragment release: BpiI | unpublished |

\* Any of these plasmids would be suitable if used as template for blunt-end cloning of basic level 0 parts (3.1.2) because the overhangs are attached with the oligonucleotides/synthesized DNA.

## 2.3 DNA oligonucleotides

Relevant oligonucleotides for the conduction of the protocol are provided in Table 2. 100 μM stocks are generated with ddH$_2$O and stored at -20°C. For PCR and Sanger sequencing reactions 10 μM working stocks are generated with ddH$_2$O and stored at -20°C.

**Tab. 2 | Oligonucleotides for the conduction of the described protocol.**

| Name | Sequence (5′-3′) | Information |
|------|------------------|-------------|
| SLo0765 | TGAAGAGCAGGCACGAACCC | Forward primer to amplify acceptor plasmid for blunt-end cloning |
| SLo0766 | AGAAGAGCGAGCACAGAGTGC | Reverse primer to amplify acceptor plasmid for blunt-end cloning |
| SLo4231 | TGCTCTTCTTTCAATAGGTCTGAGA ACGCGCGTTCTCAGACCTATTCGGT GAAGAGCT | Neutral DNA linker sequence, annealed with SLo4232 to generate double stranded DNA. |
| SLo4232 | AGCTCTTCACCGAATAGGTCTGAGA ACGCGCGTTCTCAGACCTATTGAAA GAAGAGCA | Neutral DNA linker sequence, annealed with SLo4232 to generate double stranded DNA. |
| SLo04365 | GTCGATTTTTGTGATGCTC | Primer for Sanger sequencing of level 0 parts (*see* **Note 2**). |

## 2.4 Enzymes

Any enzyme with corresponding properties can be used with their supplied buffers (*see* **Note 3**).

1. BpiI (20,000 U/mL)
2. BsaI (20,000 U/mL)
3. SapI (10,000 U/mL)
4. T4 Polynucleotide Kinase (10,000 U/mL)
5. T4 DNA Ligase (400,000 U/mL)
6. Proofreading DNA polymerase (in this case in-house purified)
7. Taq DNA polymerase (5,000 U/mL)

## 2.5 Antibiotics

All antibiotics used in this study are solved in sterile H$_2$O and are stored in 1 mL aliquots at -20°C. Stock solutions are used 1:1000 (e.g. 500 μL stock solution in 500 mL medium).

1. Spectinomycin: 120 mg/mL, stock solution
2. Ampicillin: 100 mg/mL, stock solution
3. Kanamycin: 100 mg/mL, stock solution

## 2.6 Chemicals, buffers and media components

1. Culture Dilution Buffer: 0.8% (w/v) NaCl
2. SMM: 0.3% (w/v) $KH_2PO_4$, 0.678% (w/v) $Na_2HPO_4$, 0.05% (w/v) NaCl, 0.1% (w/v) $NH_4Cl$, 1.6% (w/v) tryptone, 1.0% (w/v) yeast extract, 0.5% (v/v) glycerol
3. LB: 1% (w/v) tryptone, 0.5% (w/v) yeast extract, 1% (w/v) NaCl

## 2.7 Consumables

1. Standard petri dishes
2. Single well SBS plates (for spotting experiment)
3. 96-well microtiter plates
4. 1.5 mL reaction tubes

## 2.8 Strains

All relevant laboratory *E. coli* strains for the outlined protocol are provided in Table 3. Cells were made competent using an in-house RbCl procedure [13] (*see* **Note 4**).

**Tab. 3 | Strains used in this study.**

| Name | Relevant features | Reference |
|------|-------------------|-----------|
| *E. coli* DB3.1 (*see* **Note 5**) | F⁻*gyrA462 endA1 glnV44* Δ(*sr1-recA*) *mcrB mrr hsdS20*(r$_B^-$, m$_B^-$) *ara14 galK2 lacY1 proA2 rpsL20*(Str$^R$) *xyl5* Δ*leu mtl1* | Invitrogen |
| *E. coli* Top10 | F⁻*mcrA* Δ(*mrr-hsdRMS-mcrBC*) φ80*lacZ*ΔM15 Δ*lacX74 nupG recA1 araD139* Δ(*ara-leu*)7697 *galE15 galK16 rpsL*(Str$^R$) *endA1* λ⁻ | Invitrogen |

## 3. Methods

### 3.1 Construction and validation of basic parts

The first step in Golden Gate cloning is the design and domestication of DNA sequences into the selected standardized Golden Gate cloning hierarchy. However, before cloning can begin, the sequences must be identified and domesticated for cloning into level 0 acceptor plasmids (Fig. 2A). This procedure can be performed with different strategies as described in (3.1.1). It is important that Type IIS recognition sites used later are omitted from the basic part DNA sequences. In this protocol, basic parts are stored in level 0 plasmids. DNA sequences can be cloned into the level 0 plasmids using two different strategies. First, the blunt-end cloning strategy (3.1.2), where the fusion sites are attached to the DNA fragment to be stored as a basic part. Using this protocol will reduce the use of Type IIS enzymes, but is limited to cloning single DNA fragments. Due to the nature of blunt-end cloning, the DNA sequence can be oriented forward or reverse in the plasmid. However, release of the fragment with the appropriate Type IIS enzyme (here SapI) will result in the correct fragment with the appropriate fusion sites because the fusion sites are already added to the initial fragment (Fig. 2B). The other strategy is to use Golden Gate cloning (3.1.3) of DNA fragments in level 0, which allows the combination of multiple DNA sequences in the appropriate order, which may be necessary if Type IIS recognition sites need to be eliminated from the DNA sequence (Fig. 2C).

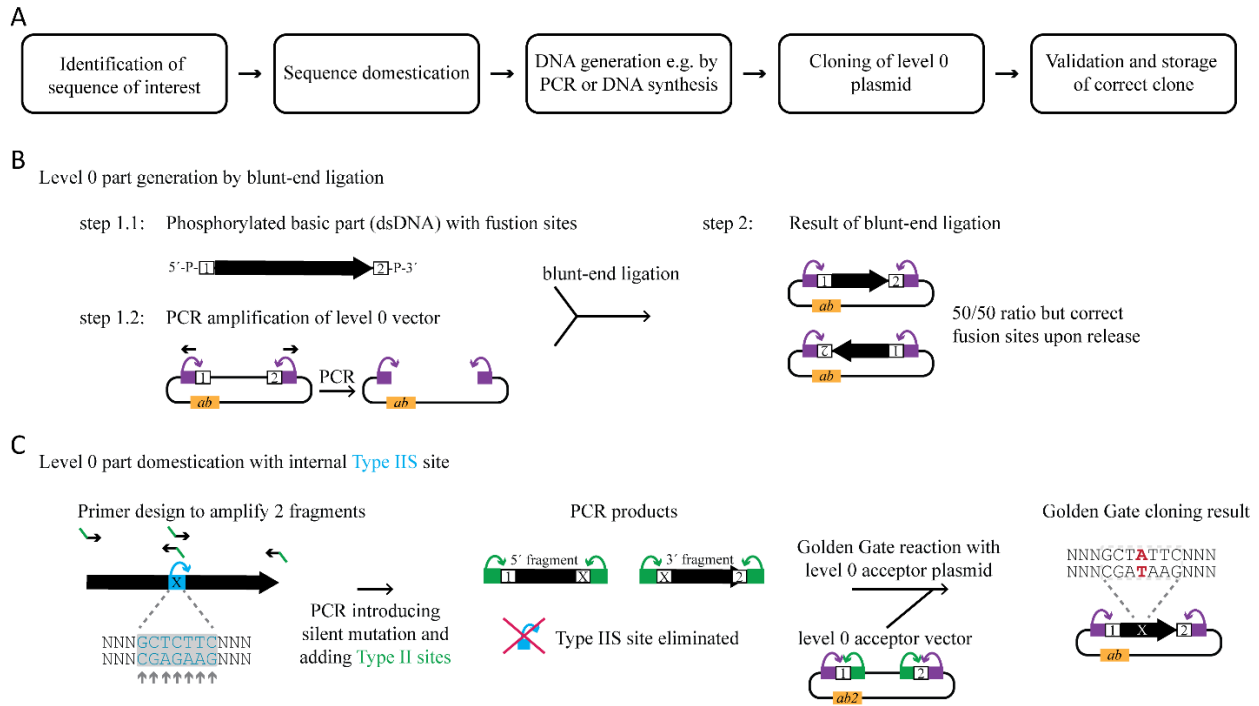

**Figure 2 | Concept and requirements of sequences domestication and basic part generation. (A)** Generalized and stepwise workflow to consider for level 0 part design, generation and validation. **(B)** Simple generation of level 0 parts by blunt-end ligation. If no Type IIS recognition site needs to be eliminated, the DNA can be amplified as a single part. The oligonucleotides must have the appropriate fusion site attached (here, 1 and 2). The DNA fragment is generated either by PCR amplification or by annealing of complementary oligonucleotides. It is important that the oligonucleotides or the PCR amplicon are phosphorylated. Each acceptor plasmid is PCR amplified without the fusion sites. Importantly the plasmid will not be phosphorylated. The DNA is mixed in a simple ligation reaction. After transformation, the fragment will be in two orientations in the plasmid. However, because the fusion sites are attached to the DNA fragment and not originate from the plasmid, both orientations are fine. **(C)** In case the DNA sequence contains one or more unwanted Type IIS recognition sites (visualized in blue) that need to be removed, amplification of multiple fragments with the appropriate Type IIS enzyme for level 0 basic part cloning (visualized in green) allows to eliminate the recognition site of any Type IIS enzyme or other small DNA sequence motif. It is important to ensure that the mutation is not deleterious, e.g. an amino acid change in the CDS. The fragments will have the appropriate fusion sites for level 0 Golden Gate cloning as well as unique fusion sites to fuse the fragments (indicated by X). The fusion sites between the fragments eliminate Type IIS recognition sites by introducing one or more sequence changes. The resulting plasmid has the recognition site eliminated (indicated by red letter in final plasmid). This strategy allows the modification of one to four consecutive nucleotides based on the typically used Type IIS enzymes with four nucleotide fusion sites.

### 3.1.1 Design of sequences (synthesis/amplification etc.)

1. For hierarchical Golden Gate cloning standards all needed sequences are generated as validated basic part plasmids, hereafter termed level 0 plasmids.

2. Familiarize with the chosen Golden Gate cloning system. In this protocol a procedure based on the Modular Cloning system of the group of Sylvestre Marillonnet is used with alterations on level 0. SapI is used to release level 0 parts to assemble transcription units avoiding assembly scars. The general steps are compatible with any other system but may require alterations of Type IIS enzymes and antibiotic selection markers throughout the steps.

3. **Important during the design process:** The part sequences must not contain any Type IIS enzyme recognition sites used in the selected Golden Gate cloning system. In the outlined case SapI (5´-GCTCTTC-3´), BpiI (5´-GAAGAC-3´) and BsaI (5´-GGTCTC-3´) and BsmBI (5´-CGTCTC-3´) must be excluded (*see* **Note 6**).

4. Individual parts can be amplified via PCR or custom synthesized by the provider of choice (*see* **Note 7**). DNA synthesis has the advantage of having full control over each nucleotide, but this may not be economic. In this protocol three fluorophores are used as an example which have been synthesized to remove unwanted enzyme recognition sites and to codon optimize the sequences. The used example promoter and terminator sequences were PCR amplified. The neutral linker DNA sequence was obtained as two oligonucleotides, subsequently annealed to generate double stranded DNA.

5. Additional to the general sequence the DNA fragments need the respective overhangs ("fusion sites") to be compatible with the selected Golden Gate cloning system which are integrated at the basic part level 0 plasmid. If a single DNA fragments are cloned in level 0 plasmid the blunt-end cloning strategy (step-by-step protocol in 3.1.2) can be used. Alternatively, or if basic parts consist of multiple fragments Golden Gate cloning for basic parts (step-by-step protocol in 3.1.3) must be used. Both strategies have different requirements for the design of the primer pairs for amplification which are outlined in step 6 and 7, respectively.

6. Amplifying basic parts for blunt ligation needs to ensure the corresponding 3 assembly position specific nucleotides are attached to the primers (depicted by 20 Ns) to ensure proper assembly of the parts for transcription units (*see* **Note 8**):

   (i) Promoter sequences
   >   forward oligo: 5´-**TTC**NNNNNNNNNNNNNNNNNNNNN-3´
   >   reverse oligo: 5´-**CAT**NNNNNNNNNNNNNNNNNNNNN-3´

   (ii) CDS sequences (*see* **Note 9**)
   >   forward oligo: 5´-**ATG**NNNNNNNNNNNNNNNNNNNNN-3´
   >   reverse oligo: 5´-**TCA**NNNNNNNNNNNNNNNNNNNNN-3´

   (iii) Terminator sequences
   >   forward oligo: 5´-**TGA**NNNNNNNNNNNNNNNNNNNNN-3´
   >   reverse oligo: 5´-**CCG**NNNNNNNNNNNNNNNNNNNNN-3´

7. Cloning basic parts by Golden Gate assembly needs to ensure the recognition sites for the used Type IIS enzyme and the respective fusion sites are added appropriately to the amplifying primers (depicted by 20 Ns) (*see* **Note 10**). Both elements are visualized in bold and underlined respectively in the examples.

(i) Promoter sequences

      forward oligo: 5´-T**GAAGAC**CA<u>TTTC</u>NNNNNNNNNNNNNNNNNNNN-3´

      reverse oligo: 5´-T**GAAGAC**CA<u>ACAT</u>NNNNNNNNNNNNNNNNNNNN-3´

(ii) CDS sequences (*see* **Note 9**)

      forward oligo: 5´-T**GAAGAC**CA<u>TATG</u>NNNNNNNNNNNNNNNNNNNN-3´

      reverse oligo: 5´-T**GAAGAC**CA<u>ATCA</u>NNNNNNNNNNNNNNNNNNNN-3´

(iii) Terminator sequences

      forward oligo: 5´-T**GAAGAC**CA<u>TTGA</u>NNNNNNNNNNNNNNNNNNNN-3´

      reverse oligo: 5´-T**GAAGAC**CA<u>ACCG</u>NNNNNNNNNNNNNNNNNNNN-3´

8. Order the designed oligonucleotides from a respective vendor and amplify the sequences of interest or directly use the synthesized fragments (*see* **Note 11**).

### 3.1.2   Construction of level 0 parts by blunt-end cloning

1. Generate and purify the basic part DNA sequence by the method of choice (e.g. PCR, DNA synthesis) according to design guidelines in section 3.1.1.

   <u>**Important:**</u> The 3 bp fusion sites must be added to the basic part as outlined in section 3.1.1 step 6.

   <u>**Important:**</u> Phosphorylated and subsequently annealed oligonucleotides with blunt ends can be used directly in step 10 (*see* **Note 12**).

2. Determine DNA quality and concentration of basic DNA parts with appropriate method(s) (e.g., gel electrophoresis, spectrophotometric, fluorescence dye assay).

3. PCR amplify the level 0 plasmid ensuring the whole plasmid is amplified, including the recognition site of the Type IIS enzyme for fragment release but excluding the fusion sites (Fig. 2B). By PCR amplification and subsequent DpnI digest high efficiency of cloning can be ensured.

4. PCR reaction mix using a proof reading DNA polymerase and reaction conditions are provided below to amplify example level 0 plasmid pSL099 using primer pair SLo765/SLo766 (*see* **Note 13**).

| Reagent | Stock concentration | Volume |
|---|---|---|
| Reaction buffer | 5 X | 10.0 µL |
| dNTPs | 10 mM | 1.0 µL |
| forward primer | 10 µM | 2.5 µL |
| reverse primer | 10 µM | 2.5 µL |
| DNA template | 1-10 ng | - |
| Polymerase | 2 U/µL | 0.5 µL |
| ddH$_2$O | - | to 50 µL |

| Step | Settings | Cycles |
|---|---|---|
| Initial denaturation | 98°C for 20 sec | - |
| PCR | 98°C for 20 sec<br>67°C for 20 sec<br>72°C for 2 min | 35 |
| Final extension | 72°C for 1 min | - |
| Hold | 12°C | - |

5. Validate PCR reaction by performing gel electrophoresis with an appropriate size standard as reference.

6. Perform DpnI digest (*see* **Note 14**) by adding the appropriate volume of DpnI reaction buffer and 0.5 µL enzyme per 50 µL PCR reaction. Digest at 37°C for at least 1 hour up (*see* **Note 15**).

7. Purify the DpnI digested sample with the method of choice and quantify the DNA concentration. It is recommended to use molarity instead of mass.

8. The fragments to be cloned into the amplified level 0 plasmid need to be phosphorylated (*see* **Note 16**). Reminder: Based on the nature of blunt-end cloning the fragments can be cloned in forward or reverse orientation. However, because the fusion sites are attached to the DNA sequence this does not matter.

9. Perform the phosphorylation according to the reaction mixture and reaction conditions outlined below (*see* **Note 17**):

| Reagent | Stock concentration | Volume |
|---|---|---|
| Double stranded DNA (or single oligonucleotides prior annealing) | 62.5 fmol/µL<br>(maximum of 6.25 pmol) | 4 µL |
| T4 DNA Ligase Buffer | 10X | 0.5 µL |
| PNK | 10 U/µL | 0.5 µL |
| ddH$_2$O | | to 5 µL |

| Step | Settings | Time |
|------|----------|------|
| Incubation | 37°C | 30 min |
| Heat inactivation | 80°C | 20 min |
| Hold | 12°C | - |

10. Because T4 Ligase is used in the phosphorylation step the DNA can just be added to the subsequent ligation step. Important: This needs to be included in the calculation for the reaction mix and the added volume is not counted towards the use of fresh T4 Ligase Buffer. The reaction mix is as following:

| Reagent | Stock concentration | Volume |
|---------|---------------------|--------|
| PCR amplified level 0 plasmid | 20 fmol/µL | 1 µL |
| Phosphorylated fragment* | 50 fmol/µL | 1 µL |
| T4 DNA Ligase Buffer | 10X | 1 µL |
| T4 DNA Ligase | 400 U/µL | 1 µL |
| ddH$_2$O | | to 10 µL |

* In case the fragment is phosphorylated as described in step 8 the used volume does not count to the total volume.

11. Perform the ligation reaction using the following settings:

| Step | Settings | Time |
|------|----------|------|
| Incubation | 16°C | overnight |
| Hold | 4°C | - |

12. Transform 2.5 µL of the reaction mix in *E. coli* strain of choice. In this example *E. coli* Top10 cells are used made competent with an in house RbCl procedure for heat shock transformation. *E. coli* Top10 is highly sensitive to the *ccdB* toxin used as an additional selection marker to prevent background. However, any *E. coli* cloning strain and protocol for competent *E. coli* cells should be appropriate.

13. Add 1 mL of LB media to cells after transformation and recover at 37°C with agitation for 30 to 60 minutes.

14. Plate transformation mixture on two plates with appropriate selection conditions (here LB spectinomycin). 100 µL of suspension is plated on the first plate. Afterwards the remaining transformation mixture is spun down 1 min at 10,000 x *g*, the supernatant is discarded, the pellet resolved in the remaining liquid and plated on the second plate.

15. Incubate plates inverted overnight at 37°C in a standing incubator.

16. Pick at least three single candidates and validate the expected insert with colony PCR or by plasmid extraction and subsequent test digest using the enzyme to release the insert.

17. Validate selected candidates using the sequencing service of choice.

18. Cryo-preserve sequence validated strain containing the level 0 basic part with a unique identifier and the relevant information for the strain (e.g., plasmid map, sequence validation raw data).

19. If enough plasmid material is left, store at -20°C for later use. If material is limited isolate plasmid with method of choice by inoculation cryo-preserved strain, determine DNA quality and concentration (*see* **Note 18**) and store until further use at -20°C.

### 3.1.3   *Construction of level 0 parts by Golden Gate cloning*

1. Purify the necessary level 0 plasmids with the method of choice, perform quality control and determine the DNA concentration. It is recommended to set the concentration of the level 0 plasmids to 50 fmol/µL.

2. PCR amplify the sequences of choice with the respective primers and their optimal reaction conditions or use synthesized DNA fragments. Keep in mind the sequence requirements outlined in 3.1.1.

3. Validate PCR amplification by gel electrophoresis with an appropriate size standard (*see* **Note 19**).

4. Purify PCR reactions after size validation with the method of choice and determine DNA concentration. It is recommended to set the DNA concentration to 50 fmol/µL to simplify the reaction conditions. This is in particular important in case many reactions are done in parallel and/or the basic part sequence consists of multiple fragments.

5. Perform Golden Gate reaction using the outlined reaction mix and reaction conditions. The example is based on the used level 0 plasmids and PCR amplified fragments (in this case the fluorophores). using a different Golden Gate system may requires to alter the used Type IIS enzyme (*see* **Note 20**).

| Reagent | Stock concentration | Volume |
|---------|--------------------|--------|
| Level 0 acceptor plasmid | 50 fmol/µL | 1 µL |
| Insert fragment (*see* **Note 21**) | 50 fmol/µL | 3 µL |
| T4 DNA Ligase Buffer | 10X | 1 µL |
| T4 DNA Ligase | 400 U/µL | 1 µL |
| BpiI | 20 U/µL | 1 µL |
| ddH$_2$O | | to 10 µL |

| Step | Settings | Cycles |
|---|---|---|
| Type IIS enzyme optimum (*see* **Note 22**) | 37°C for 3 min | 25 |
| Ligase optimum | 16°C for 4 min | |
| Final digestion step* | 50°C for 20 min | - |
| Heat inactivation | 80°C for 20 min | - |
| Hold | 12°C | - |

* At this step the ligase is inactivated which ensures minimal background because all wrongly ligated plasmids are removed from the reaction mix.

6. Transform 5 µL reaction mix of the reaction mix in *E. coli* strain of choice. In this example *E. coli* Top10 cells are used made competent with an in house RbCl procedure for heat shock transformation. *E. coli* Top10 is highly sensitive to the *ccdB* toxin used as an additional selection marker to prevent background. However, any *E. coli* cloning strain and protocol for competent *E. coli* cells should be appropriate.

7. From here follow steps 12 to 18 of section 3.1.2, the steps are identical.

### *3.2 DNA assembly of single biological parts into level 1*

Once a selection of level 0 plasmids containing the desired parts has been generated, they can be combined into TUs by Golden Gate assembly. The hierarchical Golden Gate assembly system allows for the rapid construction of modified TUs (Fig. 3). For example, if a protein is not well expressed with the initially selected promoter sequence, the user can swap to alternative sequences or even use a library-based approach to identify the optimal promoter sequence. In the outlined protocol, only one promoter is used as an example. However, the protocol could be adapted to library based cloning by simply changing the DNA sequence to a level 0 plasmid pool. The remaining steps are the same.

In this protocol, the SapI enzyme, which generates three nucleotides fusion sites, is used to allow codon-based TU cloning (Figure 3B). The use of SapI has the advantage of seamlessly assembling TUs that can subsequently be combined into higher order assemblies if appropriate level 1 plasmids are used. For simplicity, a three-part cloning is outlined in this protocol. However, the user can extend the cloning system to combine more than 3 parts by selecting appropriate fusion sites. The resulting TUs can then be assembled into higher order constructs consisting of multiple TUs as described in 3.3. Importantly, the level 1 plasmids used are compatible with level P plasmids, unlike the standard MoClo procedure where level 1 is compatible with level M (*cf.* Fig. 1C). However, level 1 plasmids that are compatible with level M are also available.

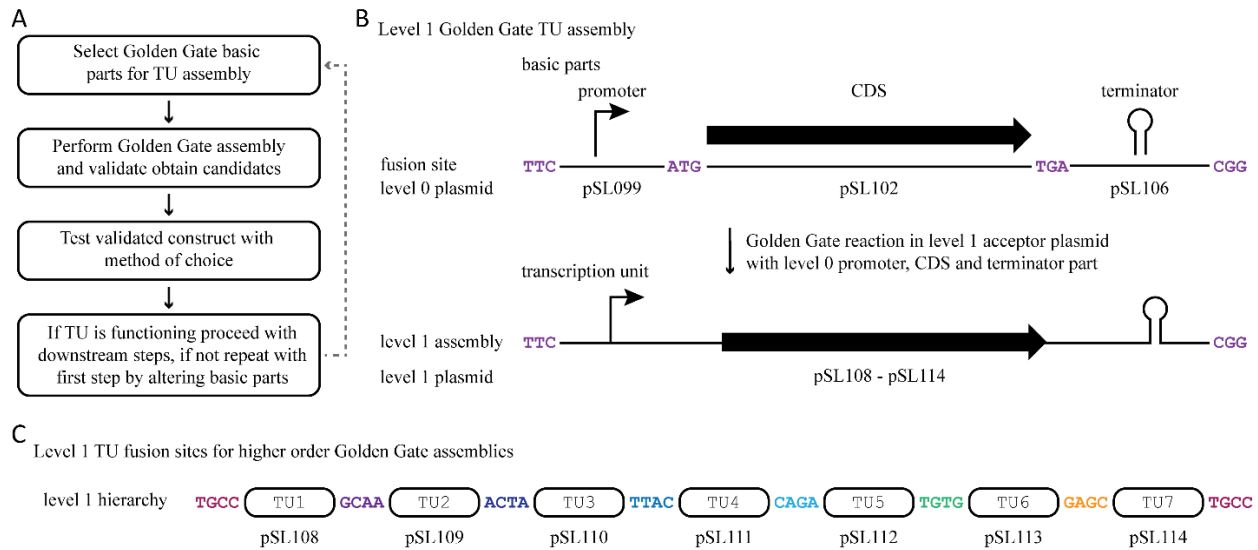

**Figure 3 | Workflow and hierarchy of level 1 Golden Gate TU assembly and subsequent assembly of multiple TUs. (A)** Generalized workflow for constructing TUs from basic parts. Functional TUs can subsequently be used for their purpose. If a TU is not functional, a variation of the TU can be generated, e.g. by changing the promoter and/or terminator (indicated by dotted line). **(B)** Golden Gate assembly using SapI-based TU cloning to use the ATG start codon and the TGA stop codon as fusion sites for scarless TU assembly. The system can be extended with additional fusion sites to allow for more complex TU assemblies, e.g. using a C- and/or N-terminal tag for fusion proteins. Using SapI allows one to choose a specific codon as a fusion site. **(C)** The Golden Gate assembled TUs can be released for the assembly of multiple TUs. There are seven positions with hierarchical fusion sites based on the original MoClo system. The system allows the Golden Gate assembly of one to six TUs in the next assembly round. The fusion sites remain the same for later stages, e.g., a multi-gene construct of TU1 to TU3 would result in the fusion sites TGCC and TTAC after release. Theoretically MoClo allows infinite rounds of DNA assembly (*cf.* Fig. 1C).

1. If not already available extract the required level 0 plasmids (here: pSLcol_16.03, pSL206, pSL209, pSL842 and pSL843) and the level 1 acceptor plasmids (here: pSL108, 109 and pSL110) with the method of choice and determine the DNA concentration. In this example three basic parts, a promoter, a coding sequence and a terminator are assembled into three different level 1 acceptor plasmids. The reason for the three different acceptor plasmids is that the transcription units can be released with the respective Type IIS enzyme and contain matching overhangs for a higher order assembly (Fig. 3C) (detailed step-by-step procedure in 3.3).

2. Set all individual plasmids to a final concentration of 50 fmol/µL.

3. In addition to the level 0 plasmid assemblies level 1 plasmids with a "neutral spacer sequence" are generated directly in the three level 1 positions, necessary for the assemblies in 3.3 (*see* **Note 23**).

4. Perform three Golden Gate reactions using the outlined reaction mix and reaction conditions. The example shown is based on three basic part positions in level 0 plasmids (Promoter pSLcol_16.03, coding sequence pSL206 [*mcherry*], pSL842 [*mTurquoise*] or

pSL843 [*gfp*], terminator pSL209) and level 1 acceptor plasmids pSL108, pSL109 or pSL110.

5. The following six Golden Gate assemblies are performed:

      mTurquoise - TU:    pSL108 with pSLcol_16.03, pSL842, and pSL209

      GFP - TU:          pSL109 with pSLcol_16.03, pSL843, and pSL209

      mCherry - TU:      pSL110 with pSLcol_16.03, pSL206, and pSL209

      Important: Neutral linker were assembled previously in pSL108 to pSL110, resulting in pSL845, pSL847 and pSL849, respectively.

| Reagent | Stock concentration | Volume |
|---|---|---|
| Level 1 acceptor plasmid | 50 fmol/µL | 1 µL |
| Each level 0 part | 50 fmol/µL | 1 µL |
| T4 DNA Ligase Buffer | 10X | 1 µL |
| T4 DNA Ligase | 400 U/µL | 1 µL |
| SapI | 10 U/µL | 1 µL |
| ddH$_2$O | | to 10 µL |

| Step | Settings | Cycles |
|---|---|---|
| Type IIS enzyme optimum (*see* **Note 22**) | 37°C for 3 min | 25 |
| Ligase optimum | 16°C for 4 min | |
| Final digestion step* | 50°C for 20 min | - |
| Heat inactivation | 80°C for 20 min | - |
| Hold | 12°C | - |

\* At this step the ligase is inactivated which ensures minimal background because all wrongly ligated plasmids are removed from the reaction mix.

6. Transform 2.5 µL reaction mix of the reaction mix in *E. coli* strain of choice. In this example *E. coli* Top10 cells are used made competent with an in house RbCl procedure for heat shock transformation. *E. coli* Top10 is highly sensitive to the *ccdB* toxin used as an additional selection marker to prevent background. However, any *E. coli* cloning strain and protocol for competent *E. coli* cells should be appropriate.

7. Add 1 mL of LB media to cells after transformation and recover at 37°C with agitation for 30 to 60 minutes.

8. Plate transformation mixture on two plates with appropriate selection conditions (here LB ampicillin). 100 µL of suspension is plated on the first plate. Afterwards the remaining transformation mixture is spun down 1 min at 10,000 x *g*, the supernatant is discarded, the pellet resolved in the remaining liquid and plated on the second plate.

9. Incubate plates inverted overnight at 37°C in a standing incubator.

10. In the outlined experiment the TU assembly can be visually validated by checking for fluorescent colonies (e.g., fluorescence binocular). In case no visible marker is used it is recommended to perform colony PCR or in case of higher order assemblies perform plasmid extraction and restriction pattern digest of the DNA with the releasing Type IIS enzyme (*see* **Note 24**).

11. Cryo-preserve generated strains according to standard procedure and document all necessary information.

12. Extract plasmid DNA, perform quality control (e.g., test digest, $A_{260/280}$ ratio) and determine concentration with method of choice. Set concentration of plasmid to 50 fmol/μL for subsequent higher order Golden Gate assembly. Store plasmid until further use at -20°C.

### 3.3 *DNA assembly of multiple TUs into higher order levels*

Using the level 1 plasmids constructed in 3.2, a higher order assembly of multiple TUs can be performed (Fig. 4). The level 1 plasmids are compatible with the MoClo level M/P cloning system developed by the group of Sylvestre Marillonnet [2]. Here a previously generated level M/P plasmid set with respective Endlinkers (EL) is used, theoretically allowing infinite rounds of DNA assembly [7,8,12]. The protocol outlined here only describes the assembly of three level 1 plasmids into multi-gene constructs. The procedure for the next round of cloning would be similar, except that some reagents would be changed, including the acceptor plasmid, Endlinker plasmid, type IIS enzyme, and selection conditions.

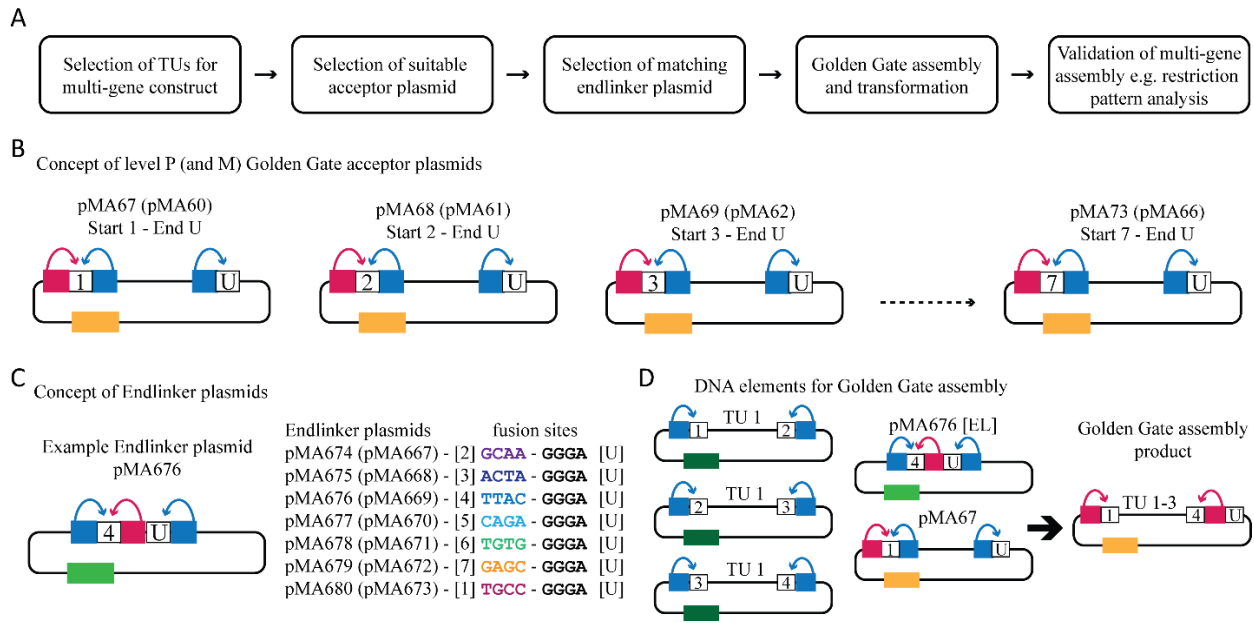

**Figure 4 | Workflow and visualization of Golden Gate multi-gene assemblies based on the MoClo system. (A)** Overview of the workflow for Golden Gate multi-gene assembly. **(B)** Concept of higher order acceptor plasmids. In contrast to level 1 plasmids (*cf*. Fig. 3), these plasmids have a universal fusion site [U] with the sequence `GGGA`. As with level 1 plasmids, there are 7 defined positions with the same fusion site sequences as in level 1. The Endlinker (EL) plasmid provides the gap-filling fragment, allowing highly flexible DNA assemblies of up to 6 TUs in a single step. **(C)** The EL provides the assembly closure fragment by releasing it with the Type IIS enzyme of the Golden Gate

reaction. In addition, the EL provides the Type IIS recognition sequence for a different Type IIS enzyme for scarless release of the multi-gene construct for additional Golden Gate assembly rounds as needed. The combination of the 7 level P (and M) plasmids with the corresponding EL plasmids provides maximum flexibility. **(D)** Example visualization of the Golden Gate assembly of three TUs (as done in this protocol) with the respective EL plasmid in the selected acceptor plasmid. The assembly results in a multi-gene construct of three TUs that can be released with a different Type IIS enzyme. Type IIS recognition sites and cleavage sites are indicated by boxes with arrows, the color indicating a switch for a different type IIS enzyme. Colored boxes in plasmids represent different selection marker genes.

1. Extract plasmids with method of choice and perform quality control and determine the concentration. For the outlined procedure the level 1 plasmids generated in 3.2, the acceptor level P plasmid pMA67 and the Endlinker plasmid pMA676 are used (*see* **Note 25**). Which results in a total of eight plasmids to purify.
2. Set the concentration of all purified plasmids to a final concentration of 50 fmol/μL.
3. Perform the following five Golden Gate reactions with the reaction mixture and conditions outlined below. Where *mTurquoise* is position 1, *gfp* is position 2 and *mCherry* is position 3 in the higher order assembly. "+" represents the respective fluorophore and "-" the neutral linker DNA.

| | |
|---|---|
| Fluorophore -/-/- | pSL844, pSL846, pSL848 in pMA67, EL pMA676 |
| Fluorophore -/+/+ | pSL844, pSL847, pSL849 in pMA67, EL pMA676 |
| Fluorophore +/-/+ | pSL845, pSL846, pSL849 in pMA67, EL pMA676 |
| Fluorophore +/+/- | pSL845, pSL847, pSL848 in pMA67, EL pMA676 |
| Fluorophore +/+/+ | pSL845, pSL847, pSL849 in pMA67, EL pMA676 |

| Reagent | Concentration | Volume |
|---|---|---|
| Level P acceptor plasmid pMA67 | 50 fmol/μL | 1 μL |
| EL plasmid pMA676 | 50 fmol/μL | 1 μL |
| Each level 1 TU position | 50 fmol/μL | 1 μL |
| T4 DNA Ligase Buffer | 10X | 1 μL |
| T4 DNA Ligase | 400 U/μL | 1 μL |
| BsaI | 20 U/μL | 1 μL |
| ddH$_2$O | | to 10 μL |

| Step | Settings | Cycles |
|---|---|---|
| Type IIS enzyme optimum (*see* **Note 22**) | 37°C for 3 min | 25 |
| Ligase optimum | 16°C for 4 min | |
| Final digestion step* | 50°C for 20 min | - |
| Heat inactivation | 80°C for 20 min | - |
| Hold | 12°C | - |

\* At this step the ligase is inactivated which ensures minimal background because all wrongly ligated plasmids are removed from the reaction mix.

4. Transform 5 µL of the Golden Gate reaction mix to the *E. coli* strain of choice. In this procedure in house prepared RbCl chemical competent *E. coli* Top10 cells are used.

5. Select after transformation on the appropriate medium, in this case LB kanamycin at 37°C overnight.

6. Validate single colonies with the method of choice for the correct assembled plasmid. In case of larger sizes plasmid extraction and test restriction digest analysis with the assembly release Type IIS enzyme is recommended. Colony PCR becomes inefficient with larger assemblies.

7. Validated strains should be cryo-preserved with unique identifiers and all necessary data needs to be stored accordingly.

8. Assembly can in theory have unlimited number of rounds using the level M/P plasmids in consecutive steps (Fig. 1C). However, in this protocol the higher order assembly of three level 1 TUs is the final step and the assemblies are compared in in step 3.4.

### 3.4  *Macroscopic validation of assembled multi-gene constructs*

Once DNA constructs are generated, they need to be analyzed to study their biology or application potential. These experiments are very diverse and need to be adapted to the specific question. In this protocol, several multi-gene constructs containing up to three highly expressed fluorescent proteins were generated. To test their expression and potential effect on bacterial growth, a spotting experiment was performed to compare the different strains (Fig. 5). This procedure was used for visualization purposes and could be replaced by other methods such as plate reader or flow cytometry based monitoring. However, this part is intended as an example and must be adapted to the assembled multi-gene construct of interest.

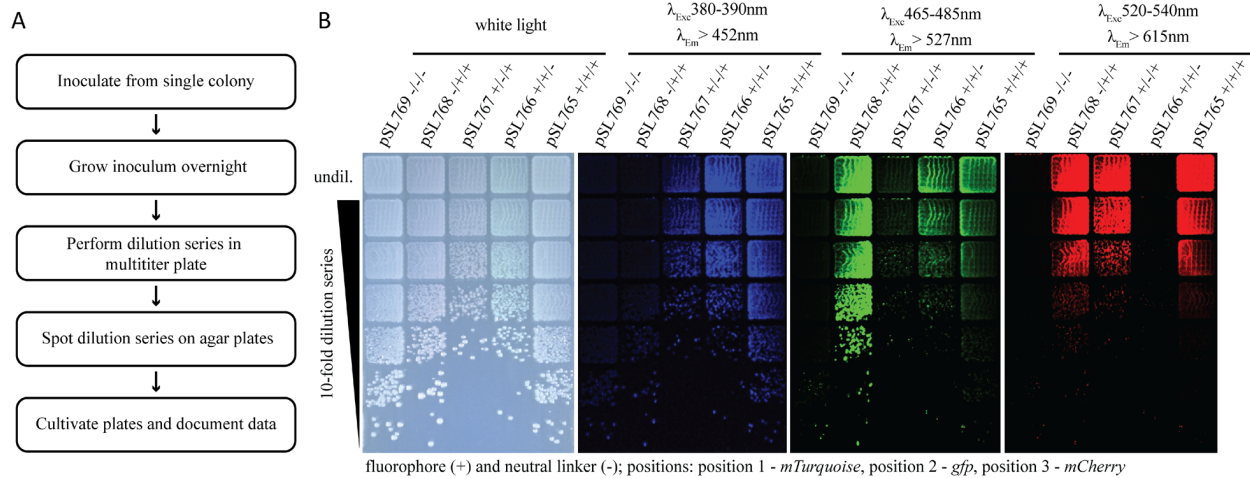

fluorophore (+) and neutral linker (-); positions: position 1 - *mTurquoise*, position 2 - *gfp*, position 3 - *mCherry*

**Figure 5 | Result of higher order assembly of multiple TUs. (A)** Visualization of experimental workflow for growth assay in the form of a spotting assay. **(B)** Results of spotted cultures of strains with the assembled level P multi-gene constructs. Shown are images of the same plate under white light and the respective fluorophore specific conditions. Fluorescence corresponds to the expected expression of fluorophores from the assembled multi-gene constructs.

1. Streak the cryo-preserved strains on appropriate selective medium and incubate overnight at 37°C.
2. Inoculate from a single colony for each strain into a glass tube with 3 mL of LB kanamycin. Cultivate shaking with 180 rpm at 37°C overnight or until all strains reach the stationary phase.
3. Prepare single well plates with appropriate selective medium for spotting experiment (*see* **Note 26**).
4. Transfer 200 µL dense culture of the strains of interest into either column 1 or row A depending on if less or more than 8 strains are analyzed.
5. Add appropriate volume of culture dilution buffer to the remaining wells. The volume depends on the dilution series e.g., for 10-fold dilutions dispense 180 µL, for 5-fold dilutions dispense 160 µL.
6. Transfer the appropriate volume of the dense culture to the first dilution step using either an 8- or 12-channel pipette. Mix dilutions properly by pipetting up and down multiple times. Discard pipette tips and perform next dilution step by aspirating the just generated dilution with fresh sterile tips. Repeat for the appropriate number of dilution steps.
7. Spot the dilution series onto the single well plate using either a multi channel pipette (*see* **Note 27, 28**) or if available a pin-pad based screening robot (Fig. 5) (*see* **Note 29**).
8. Let liquid of spots evaporate in a sterile environment before placing the lid and incubate plates overnight at 37°C.
9. If intense colors are needed the plates can be incubated in the fridge for 6 hours to accumulate proteins before documentation with an appropriate imaging system.
10. Colony sizes may provide information of potential fitness defects in case the diameter is significantly smaller compared to the control strain.

## 3.5    Summary & Perspectives

This step-by-step protocol describes how to design and generate level 0 parts for TU assembly with three nucleotide fusion sites, allowing codon-based cloning and avoiding unwanted sequences at the TU level. The core elements of this protocol are outlined in sections 3.1 and 3.2, but an example of its compatibility with the Modular Cloning standard developed by Sylvestre Marillonnet is given in section 3.3. A brief example of testing higher order DNA assemblies on solid media is given in 3.4. Fluorescent genes are used as an example, but the procedure could be adapted to any other type of multi-fragment assembly, such as the expression of heterologous pathways. In particular, the procedure outlines the onboarding of new level 0 parts, which allows the creation of sequence libraries to test different combinations to identify the best performing DNA construct.

## 4. Notes

1.  This protocol uses throughout an open source procedure magnetic bead plasmid extraction and DNA purification procedure [14].
2.  This primer is sufficient for sequencing smaller basic parts. If larger fragments are domesticated as basic parts, it may be necessary to use additional primer(s) to sequence the entire insert. For very large constructs, it may be more economical to use a whole plasmid sequencing service.
3.  All enzymes used in this protocol were supplied by NEB. However, any other supplier of enzymes should be fine.
4.  Any type of competent cells can be used. However, it is recommended to test their competence in advance to ensure high cloning efficiency.
5.  This strain is relevant for the propagation of *ccdB* containing plasmids. However, any other *ccdB* resistant strain would be suitable. *ccdB* is a toxin that inhibits gyrase [15]. The strain must contain the *ccdA* antitoxin or the appropriate gyrase mutation(s).
6.  Ensure that no Type IIS recognition sites are formed in subsequent assembly steps and that the acceptor plasmids do not contain the Type IIS recognition sites used, except where necessary for DNA assembly purposes.
7.  The advantage of DNA synthesis is that the sequence is free in the design space. This allows, for example, to easily eliminate DNA motifs and/or to freely design the codon usage of a CDS. A good tool for automated, rational *in silico* DNA sequence optimization is DNA Chisel [16].
8.  The nucleotides are specific for the plasmid set used here. If other Golden Gate systems are used, the fusion site sequences may differ. If custom fusion sites are used, it is recommended to use the appropriate tools to predict ideal fusion site combinations (e.g. the NEB tools driven by ligase fidelity data[11,17,18].

9. Ensure that the start and stop codon are not duplicated due to the ATG and TGA selected as the fusion site for CDS primers in this Golden Gate cloning system.

10. For the plasmid set used here, Golden Gate cloning in level 0 uses BpiI. This may vary depending on the Golden Gate cloning plasmids used.

11. If gene synthesis is performed instead of fragment synthesis, vendors offer subcloning into custom plasmids. In this case, it is useful to deposit the level 0 plasmid used with the vendor to directly obtain validated level 0 parts. In this case, the protocol steps for level 0 plasmid generation can be skipped and the user can proceed directly to step 3.2.

12. Due to the PNK heat inactivation step, the phosphorylated oligonucleotides are annealed after the PNK reaction by briefly boiling the mixture and slowly cooling the mixture to room temperature. A PCR cycler is used for this. The program heats the mixture to 95°C for 5 minutes and then cools the samples to 12°C at the minimum ramp speed of the PCR cycler (here 0.1°C/sec). After several freeze-thaw cycles, it is recommended to repeat the annealing reaction.

13. This PCR amplification strategy and the attachment of the fusion site to the DNA fragments allows great flexibility with the fusion sites and allows modifications to be made as needed. For example, the use of more complex cloning schemes that require the addition of C- and/or N-terminal fusion tags to the gene of interest. When using alternative fusion sites, it is recommended to predict optimal fusion sites using appropriate computational tools, e.g. the NEB tools which are driven by ligase fidelity data [11,17,18].

14. DpnI digests methylated GATC sequence motifs on the adenin base. GATC is methylated by the Dam methylase in standard *E. coli* strains. The motif randomly occurs every 256 bp. If an *E. coli* Δ*dam* strain is used, this reaction is not effective. However, this step is highly recommended to reduce potential background during subsequent cloning.

15. DpnI digestion can also be performed overnight and would reduce the amount of enzyme used to 0.1 µL per reaction.

16. To avoid this step, oligonucleotides and/or DNA fragments can be ordered already phosphorylated. Alternatively, oligonucleotides can be phosphorylated prior to PCR, similar to the steps described in step 8. However, it is recommended to check the reaction conditions and units of T4 polynucleotide kinase used.

17. If multiple fragments are phosphorylated, it is recommended to make a 2.5 x master mix containing water, T4 DNA Ligase Buffer and T4 PNK.

18. It is recommended to store the validated level 0 parts with a defined molarity, so that they can be used directly in subsequent Golden Gate assemblies without the need for frequent volume calculations to match the required mass.

19. If the sequence was amplified from a plasmid containing the same antibiotic resistance marker as the level 0 plasmid, a careful performed DpnI digestion is recommended (see step 5 in 3.1.2).

20. In contrast to 3.1.2, the fusion sites are already present in the level 0 plasmid. The fusion sites of the plasmid and the fragment(s) must match to ensure an efficient reaction and to use the fragment(s) for transcription unit assembly.
21. In case of a single fragment Golden Gate cloning reaction, the fragment can be provided in ~ 3-fold access to the acceptor plasmid. However, for multiple fragments, equimolar amounts of all DNA fragments are recommended.
22. If alternative Type IIS enzymes are used, it is imperative to check the compatibility of the enzyme with the protocol. For example, BsmBI has an optimum temperature of 55°C. However, this temperature already inactivates the ligase. Therefore, the temperature would be increased to 42°C to have the optimum for BsmBI activity but minimal inactivation of the ligase.
23. Neutral spacer sequences were generated using *SiteOut* [19]. The DNA was ordered as oligonucleotides, phosphorylated, annealed and inserted into all PCR amplified level 1 acceptor plasmids using the standard blunt-end cloning procedure.
24. Validation by sequencing is not necessary in most cases because the basic parts have not been amplified by PCR and have been sequence validated at the level 0. However, if the construct causes high load or toxic products, it may mutate. An indication of this could be a low number of colonies obtained or very low efficiency during restriction pattern analysis of generated constructs. If the sequence cannot be altered in this case, it may be necessary to use an alternative host as a recipient for the Golden Gate assembly in order to propagate the DNA.
25. In the original MoClo system, level 1 TUs are assembled in level M. In this protocol, we use level 1 plasmids that are compatible with level P. However, level 1 plasmids for SapI TU assembly are available for assembly in level M.
26. Ensure that the plates are dry but not overdried for best quality results. The surface should be flat.
27. The volume to be dispensed depends on the medium and must be tested beforehand to avoid merging of individual spots.
28. To improve manual spotting coordination, it may be helpful to place a grid under the single well plate, such as the lid of a 96-well microplate or the inlet of a 96-tip tip box.
29. In this protocol, the Singer Instruments Rotor HDA+, a screening robot with reusable pin pads and a 7x7 spotting program, was used to reliably spot squares on agar plates.

## Acknowledgments

This work was supported by the Max Planck Society within the framework of the MaxGENESYS project (DS) and the European Union (NextGenerationEU) via the European Regional Development Fund (ERDF) by the state Hesse within the project "biotechnological production of reactive peptides from waste streams as lead structures for drug development" (DS). We are grateful to all laboratory members for extensive discussions on Golden Gate cloning in particular Tania S. Köbel for her superior technical support. We thank Torsten Waldminghaus for providing plasmids. All material is available from the corresponding author upon request.